\documentclass{rspublic}
\usepackage{graphicx}
\usepackage{epsfig}
\usepackage{subfigure}

\def\spose#1{\hbox to 0pt{#1\hss}}

\def\kms{\ifmmode {\rm\,km\,s^{-1}}\else ${\rm\,km\,s^{-1}}$\fi}

\def\kmsmpc{\ifmmode {\rm\,km\,s^{-1}\,Mpc^{-1}}\else
${\rm\,km\,s^{-1}\,Mpc^{-1 }}$\fi}
\def\hmpcs{{\rm\thinspace $h^{-1}\rm\thinspace Mpc$\ }} 
\def\hmpc{{\rm\thinspace $h^{-1}\rm\thinspace Mpc$}} 
\def\Msun{{\rm\,M_\odot}}
\def\msun{{\rm\,$h^{-1} M_\odot$}}

\def\ergps{\ifmmode {\rm\,erg\,s^{-1}}\else ${\rm\,erg\,s^{-1}}$\fi}
\def\ergpscm2{\ifmmode {\rm\,erg\,s^{-1}\,cm^{-2}}\else
    ${\rm\,erg\,s^{-1}\,cm^{-2}}$\fi}

\def\deg{\ifmmode {^{\circ}}\else {$^\circ$}\fi}
\def\degr{\ifmmode {^{\circ}}\else {$^\circ$}\fi}
\def\degs{\ifmmode {^{\circ}}\else {$^\circ$}\fi}

\def\etal{{\it et al.~}}
\def\h3Mpc{h^{-3}{\rm Mpc}^3}

\def\arcsec{\ifmmode {^{\prime\prime}}\else $^{\prime\prime}$\fi}
\def\asec{\ifmmode {^{\prime\prime}}\else $^{\prime\prime}$\fi}
\def\arcmin{\ifmmode {^{\prime}}\else $^{\prime}$\fi}
\def\amin{\ifmmode {^{\prime}}\else $^{\prime}$\fi}

\def\secper{\ifmmode \rlap.{^{s}}\else $\rlap{.}{^{s}} $\fi}
\def\minper{\ifmmode \rlap.{^{m}}\else $\rlap{.}{^m} $\fi}
\def\secspt{\ifmmode \rlap.{^{\prime\prime}}\else
    $\rlap.{^{\prime\prime}}$\fi}
\def\arcsper{\ifmmode \rlap.{^{\prime\prime}}\else
    $\rlap.{^{\prime\prime}}$\fi}
\def\minspt{\ifmmode \rlap.{^{\prime}}\else
    $\rlap.{^{\prime}}$\fi}
\def\arcmper{\ifmmode \rlap.{^{\prime}}\else
    $\rlap.{^{\prime}}$\fi}
\def\spose#1{\hbox to 0pt{#1\hss}}
\def\simlt{\mathrel{\spose{\lower 3pt\hbox{$\mathchar"218$}}
     \raise 2.0pt\hbox{$\mathchar"13C$}}}
\def\simgt{\mathrel{\spose{\lower 3pt\hbox{$\mathchar"218$}}
     \raise 2.0pt\hbox{$\mathchar"13E$}}}

\def\refindent{\par\noindent\parskip=2pt\hangindent=3pc\hangafter=1 }

\def\book#1;#2;#3 {\refindent{#1, }{in {\it{#2},} }{#3}}

\def\U300{\ifmmode{U_{300}}\else{$U_{300}$}\fi}
\def\B450{\ifmmode{B_{450}}\else{$B_{450}$}\fi}
\def\V606{\ifmmode{V_{606}}\else{$V_{606}$}\fi}
\def\I814{\ifmmode{I_{814}}\else{$I_{814}$}\fi}
\def\J110{\ifmmode{J_{110}}\else{$J_{110}$}\fi}
\def\H160{\ifmmode{H_{160}}\else{$H_{160}$}\fi}

\begin{document} 
\title[Simulating the Universe] {Simulating the formation of cosmic
structure}
\author[C. S. Frenk]{C. S. Frenk} 
\affiliation{Physics Department, University of Durham, Durham DH1 3LE,
England }  
\label{firstpage}
\maketitle

\begin{abstract}{Cosmic structure, dark matter, gas dynamics, galaxy
formation, computer simulations} A timely combination of new theoretical
ideas and observational discoveries has brought about significant advances
in our understanding of cosmic evolution. Computer simulations have played
a key role in these developments by providing the means to interpret
astronomical data in the context of physical and cosmological theory. In
the current paradigm, our Universe has a flat geometry, is undergoing
accelerated expansion and is gravitationaly dominated by elementary
particles that make up cold dark matter. Within this framework, it is
possible to simulate in a computer the emergence of galaxies and other
structures from small quantum fluctuations imprinted during an epoch of
inflationary expansion shortly after the Big Bang. The simulations must
take into account the evolution of the dark matter as well as the gaseous
processes involved in the formation of stars and other visible
components. Although many unresolved questions remain, a coherent picture
for the formation of cosmic structure in now beginning to emerge.
\end{abstract}

\section{Introduction} 

The origin of structure in the Universe is a central problem in
Physics. Its solution will not only inform our understanding of the
processes by which matter became organized into galaxies and clusters,
but it will also help uncover the identity of the dark matter, offer
insights into events that happened in the early stages of the Big Bang
and provide a useful check on the values of the fundamental
cosmological parameters estimated by other means.

Because of its non-linear character, lack of symmetry and general
complexity, the formation of cosmic structure is best approached
theoretically using numerical simulations. The problem is well posed
because the initial conditions -- small perturbations in the density
and velocity field of matter -- are, in principle, known from Big Bang
theory and observations of the early Universe, while the basic
physical principles involved are understood. The behaviour of the dark
matter is governed primarily by gravity, while the formation of the
visible parts of galaxies involves gas dynamics and radiative
processes of various kinds.  Using cosmological simulations it is possible 
to follow the development of structure from primordial pertubations to
the point where the model can be compared with observations.

Over the past few years, there has been huge progress in quantifying
observationally the properties of galaxies not only in the nearby universe,
but also in the very distant universe. Since the clustering
pattern of galaxies is rich with information about physics and cosmology,
much effort is invested in mapping the distribution of galaxies at
different epochs. Two large ongoing surveys, the US-based Sloan Digital Sky
Survey (York \etal 2000), and the Anglo-Australian ``2-degree field galaxy
redshift survey'' (2dFGRS, Colless \etal 2001), are revolutionizing our
view of the nearby universe with order of magnitude increases in the
amount of available data. Similarly, new data collected in the past five
years or so have, for the first time, opened up the high redshift
universe\footnote{In cosmology, distances to galaxies are estimated from
the redshift of their spectral lines; higher redshifts correspond to more
distant galaxies and thus to earlier epochs.} to detailed statistical study
(Steidel \etal 1996).

The advent of large computers, particularly parallel supercomputers,
together with the development of efficient algorithms, has enabled the
accuracy and realism of simulations to keep pace with observational
progress. With the wealth of data now available, simulations are
essential to interpret astronomical data and to link them to 
physical and cosmological theory.

\section{Building a model} 

To build a model of large-scale structure, four key ingredients need
to be specified: (i) the content of Universe, (ii) the initial
conditions, (iii) the growth mechanism, and (iv) the values of
fundamental cosmological parameters. I now discuss each of these in
turn.

\subsection{The content of the Universe} 

Densities are usually expressed in terms of the cosmological density
parameter, $\Omega=\rho/\rho_{crit}$, where the critical density,
$\rho_{crit}$, is the value that makes the geometry of the Universe
flat. The main constituents of the Universe and their contribution to
$\Omega$ are listed in Table~1.

\bigskip
\begin{figure}[h]
\centerline{Table 1. {\it The content of the Universe}}
\medskip
\begin{center}
\begin{tabular}{ll}
\hline \textbf{Component} & \textbf{Contribution to $\Omega$} \\ \hline

CMB radiation &  $\Omega_r=4.7\times 10^{-5}$ \\

massless neutrinos &  $\Omega_\nu=3\times 10^{-5}$ \\

massive neutrinos &  
$\Omega_\nu=6\times 10^{-2}{\left(<m_\nu>\over 1 {\rm ev}\right)}$ \\ 
baryons & $\Omega_b=0.037 \pm 0.009$ \\

(of which stars  ) & 
$\Omega_s=(0.0023 - 0.0041)\pm0.0004$ \\ 

dark matter & $\Omega_{\rm dm}\simeq 0.3$\\


dark energy  &  $\Omega_{\Lambda}\simeq 0.7$   \\
\hline
\end{tabular}
\end{center}
\end{figure} 

The main contribution to the extragalactic radiation field today is
the cosmic microwave background (CMB), the redshifted radiation left
over from the Big Bang. These photons have been propagating freely
since the epoch of ``recombination", approximately 300,000 years after
the Big Bang. The CMB provides a direct observational window to the
conditions that prevailed in the early Universe. The Big Bang also
produced neutrinos which today have an abundance comparable to that of
photons. We do not yet know for certain what, if any, is the mass of
the neutrino, but even for the largest masses that seem plausible at
present, $\sim 0.1$eV, neutrinos make a negligible contribution to the
total mass budget (although they could be as important as
baryons). The abundance of baryons is now known with reasonable
precision from comparing the abundance of deuterium predicted by Big
Bang theory with observations of the absorption lines produced by
intergalactic gas clouds at high redshift seen along the line-of-sight
to quasars (Tytler \etal 2000).  Baryons, the overwhelming majority
of which are {\it not} in stars today, are also dynamically
unimportant (except, perhaps, in the cores of galaxies). 

Dark matter makes up most of the matter content of the Universe today. To
the now firm dynamical evidence for its existence in galaxy halos, even
more direct evidence has been added by the phenomenon of gravitational
lensing which has now been detected around galaxy halos (e.g. Fischer \etal
2000, McKay \etal 2001, Wilson \etal 2001), in galaxy clusters (e.g. Clowe
\etal 2000), and in the general mass field (e.g. Van Waerbeke \etal 2001
and references therein). The distribution of dark matter in rich clusters
can be reconstructed in fair detail from the weak lensing of distant
background galaxies in what amounts virtually to imaging the cluster dark
matter. Various dynamical tests are converging on a value of $\Omega_{\rm
dm}\simeq0.3$, which is also consistent with independent determinations
such as those based on the baryon fraction in clusters (White \etal 1993,
Evrard 1997), and on the evolution in the abundance of galaxy clusters (Eke
\etal 1998, Borgani \etal 2001). Since $\Omega_{\rm dm}$ is much larger than 
$\Omega_b$, it follows that the dark matter cannot be made of baryons. The
most popular candidate for the dark matter is a hypothetical elementary
particle like those predicted by supersymmetric theories of particle
physics. These particles are referred to generically as cold dark matter or
CDM. (Hot dark matter is also possible, for example, if the neutrino had a
mass of $\sim5$ eV. However, early cosmological simulations showed that the
galaxy distribution in a universe dominated by hot dark matter would not
resemble that observed in our Universe (White, Frenk and Davis 1983).)

A recent addition to the cosmic budget is the dark energy, direct evidence
for which was first provided by studies of type Ia supernovae (Riess \etal
1998, Perlmutter \etal 1999){\footnote{The possibility that dark energy
might be the dynamically dominant component had been anticipated by
theorists from studies of the cosmic large-scale structure (see
e.g. Efstathiou \etal 1990), and was considered in the first simulations of
structure formation in cold dark matter universes (Davis \etal
1985).}. These presumed `standard candles' can now be observed at redshifts
between 0.5 and 1 and beyond. The more distant ones are fainter than would
be expected if the universal expansion were decelerating today, indicating
that the expansion is, in fact, accelerating. Within the standard Friedmann
cosmology, there is only one agent that can produce an accelerating
expansion. This is nowadays known as dark energy, a generalization of the
cosmological constant first introduced by Einstein, which could, in
principle, vary with time. The supernova evidence is consistent with the
value $\Omega_{\Lambda}\simeq 0.7$.  Further, independent evidence for dark
energy is provided by a recent joint analysis of CMB data (see next
section) and the 2dFGRS (Efstathiou \etal 2002).

Amazingly, when all the components are added together, the data are
consistent with a flat universe: 

\begin{equation}
\Omega=\Omega_b+\Omega_{\rm dm}+\Omega_\Lambda\simeq 1 
\label{eq:omega}
\end{equation} 

\subsection{The initial conditions}

The idea that galaxies and other cosmic structures are the result of the
slow amplification by the force of gravity of small primordial
perturbations present in the mass density at early times goes back, at
least, to the 1940s (Lifshitz 1946). However, it was only in the early
1980s that a physical mechanism capable of producing small perturbations
was identified. This is the mechanism of inflation, an idea due to Guth
(1981), which changed the face of modern cosmology. Inflation is produced
by the dominant presence of a quantum scalar field which rolls slowly from
a false to the true vacuum, maintaining its energy density approximately
constant and causing the early Universe to expand exponentially for a brief
period of time. Quantum fluctuations in the inflaton field are blown up to
macroscopic scales and become established as genuine adiabatic ripples in
the energy density. Simple models of inflation predict the general
properties of the resulting fluctuation field: it has Gaussian distributed
amplitudes and a near scale-invariant power spectrum (Starobinskii 1982).

After three decades of ever more sensitive searches, evidence for the
presence of small fluctuations in the early universe was finally
obtained in 1992. Since prior to recombination the matter and
radiation fields were coupled, fluctuations in the mass density are
reflected in the temperature of the radiation. Temperature
fluctuations in the CMB were discovered by the COBE satellite (Smoot
\etal 1992) and are now being measured with ever increasing 
accuracy, particularly by detectors deployed in long-flight balloons
(de Bernardis \etal 2000, Hanany \etal 2000, Leitch \etal 2002).  The
spectrum of temperature fluctuations is just what inflation predicts:
it is scale invariant on large scales and shows a series of
``Doppler'' or ``acoustic" peaks which are the result of coherent
acoustic oscillations experienced by the photon-baryon fluid before
recombination. The characteristics of these peaks depend on the values
of the cosmological parameters. For example, the location of the first
peak is primarily determined by the large-scale geometry of the
Universe and thus by the value of $\Omega$. Current data imply a flat
geometry, consistent with eqn.~\ref{eq:omega}.

The spectrum of primordial fluctuations generated, for example, by
inflation evolves with time in a manner that depends on the content of
the Universe and the values of the cosmological parameters. The dark
matter acts as a sort of filter, inhibiting the growth of certain
wavelengths and promoting the growth of others. Following the
classical work of Bardeen \etal (1986), transfer functions for
different kinds of dark matter (and different types of primordial
fluctuation fields, including non-Gaussian cases) have been
computed. In Gaussian models, the product (in Fourier space) of the
primordial spectrum and the transfer function, together with the
growing mode of the associated velocity field, provides the initial
conditions for the formation of cosmic structure.

\subsection{Growth mechanism} 

Primordial fluctuations grow by gravitational instability: overdense
fluctuations expand linearly, at a retarded rate relative to the
Universe as a whole, until eventually they reach a maximum size and
collapse non-linearly to form an equilibrium (or `virialized') object
whose radius is approximately half the physical size of the
perturbation at maximum expansion. The theory of fluctuation growth is
lucidly explained by Peebles (1980).

Although gravitational instability is now widely accepted as the primary
growth mechanism responsible for the formation of structure, it is only
very recently that firm empirical evidence for this process was
found. Gravitational instability causes inflow of material around overdense
regions.  From the perspective of a distant observer, this flow gives rise
to a characteristic infall pattern which is, in principle, measurable in a
galaxy redshift survey by comparing the two-point galaxy correlation
function along and perpendicular to the line-of-sight. In this space, the
infall pattern resembles a butterfly (Kaiser 1987).  This pattern has been
clearly seen for the first time in the 2dFGRS (Peacock \etal
2001)\footnote{Strictly speaking the `butterfly' pattern does not prove the
existence of infall since the continuity equation would ensure a similar
pattern even if velocities were induced by non-gravitational
processes. However, it can be shown that such velocities, if present, would
rapidly decay.}.

\subsection{Cosmological parameters} 

After decades of debate, the values of the fundamental cosmological
parameters are finally being measured with some degree of precision. The
main reason for this is the accurate measurement of the acoustic peaks in
the CMB temperature anisotropy spectrum whose location, height and shape
depend on the values of the cosmological parameters. Some parameter
degeneracies exist but some of these can be broken using other data, for
example, the distant Type Ia supernovae or the 2dFGRS (eg. Efstathiou \etal
2002). The CMB data alone do not constrain the Hubble constant, but there
is a growing consensus from the HST key project (Freedman \etal 2001), and
other methods, that its value, in units of 100 km s$^{-1}$ Mpc,$^{-1}$ is
$h=0.70 \pm 0.07$. In addition to $h$ and the other parameters listed in
Table~1, the other important number in studies of large-scale structure is
the amplitude of primordial density fluctuations which is usually
parametrized by the quantity $\sigma_8$ (the linearly extrapolated value of
the top-hat filtered fluctuation amplitude on the fiducial scale of
8\hmpc). The best estimate of this quantity comes from the observed
abundance of rich galaxy clusters which gives $\sigma_8\Omega^{0.6}=0.5$,
with an uncertainty of about 10\% (Eke, Cole \& Frenk 1996, Viana \& Liddle
1996, Pierpaoli \etal 2001).

\section{Cosmological simulations}

Operationally, the problem of the cosmic large-scale structure can be
divided into two parts: understanding the clustering evolution of the dark
matter and understanding the gaseous and radiative processes that lead to
the formation of galaxies. Specialized simulation techniques have been
developed to tackle both aspects of the problem. The evolution of the dark
matter is most often calculated using N-body techniques, implemented
through a variety of efficient algorithms, such as P$^3$M
(Particle-particle/particle-mesh; Efstathiou \etal 1985), AP$^3$M (the
adaptive mesh version of P$^3$M; Couchman \etal 1995) and hierarchical
trees (Barnes \& Hut 1986, Springel \etal 2001, Stadel 2000). Gaseous and
radiative processes are followed by combining a hydrodynamics code with an
N-body code. Numerical hydrodynamic techniques used in cosmology include
Eulerian methods (Cen 1992), Lagrangian codes based on Smooth Particle
Hydrodynamics (SPH) (Gingold \& Monaghan 1977), and hybrid codes
(e.g. Gnedin 1995, Pen 1998). These techniques have different strengths and
weaknesses, but they all give similar results in the simplest cosmological
problems where a detailed comparison has been performed (Frenk \etal 1999).

There has been a rapid growth in the size and power of cosmological
simulations in the two and a half decades since this technique was
introduced into the subject by Peebles (1970). One way to measure this
growth is by the number of particles employed in the simulations. The
size of the largest simulations has grown exponentially, in a manner
reminiscent of the well-known ``Moore's law'' that describes the
increase in cpu speed with time, except that the advent of massively
parallel supercomputers led to a sudden order-of-magnitude jump in
size towards the end of the past decade.  The largest simulations
carried out to date are the 1-billion particle ``Hubble volume,''
N-body simulations performed by the Virgo consortium, an international
collaboration of reseachers in the UK, Germany and Canada.

\subsection{Large-scale structure} 

Figure~\ref{fig:4panel} illustrates the spatial distribution of {\it dark
matter} at the present day, in a series of simulations covering a large
range of scales. Each panel is a thin slice of the cubical simulation
volume and shows the slightly smoothed density field defined by the dark
matter particles. In all cases, the simulations pertain to the 
``$\Lambda$CDM" cosmology, a flat cold dark matter model in which
$\Omega_{\rm dm}=0.3$, $\Omega_{\Lambda}=0.7$ and $h=0.7$.  The top-left
panel illustrates the Hubble volume simulation: on these large scales, the
distribution is very smooth. To reveal more interesting structure, the top
right panel displays the dark matter distribution in a slice from a volume
approximately 2000 times smaller. At this resolution, the characteristic
filamentary appearance of the dark matter distribution is clearly
visible. In the bottom-right panel, we zoom again, this time by a factor of
5.7 in volume. We can now see individual galactic-size halos which
preferentially occur along the filaments, at the intersection of which
large halos form that will host galaxy clusters.  Finally, the bottom-left
panel zooms into an individual galactic-size halo. This shows a large
number of small substructures that survive the collapse of the halo and
make up about 10\% of the total mass (Klypin \etal 1999, Moore \etal 1999)

\begin{figure}
\centerline{\epsfig{file=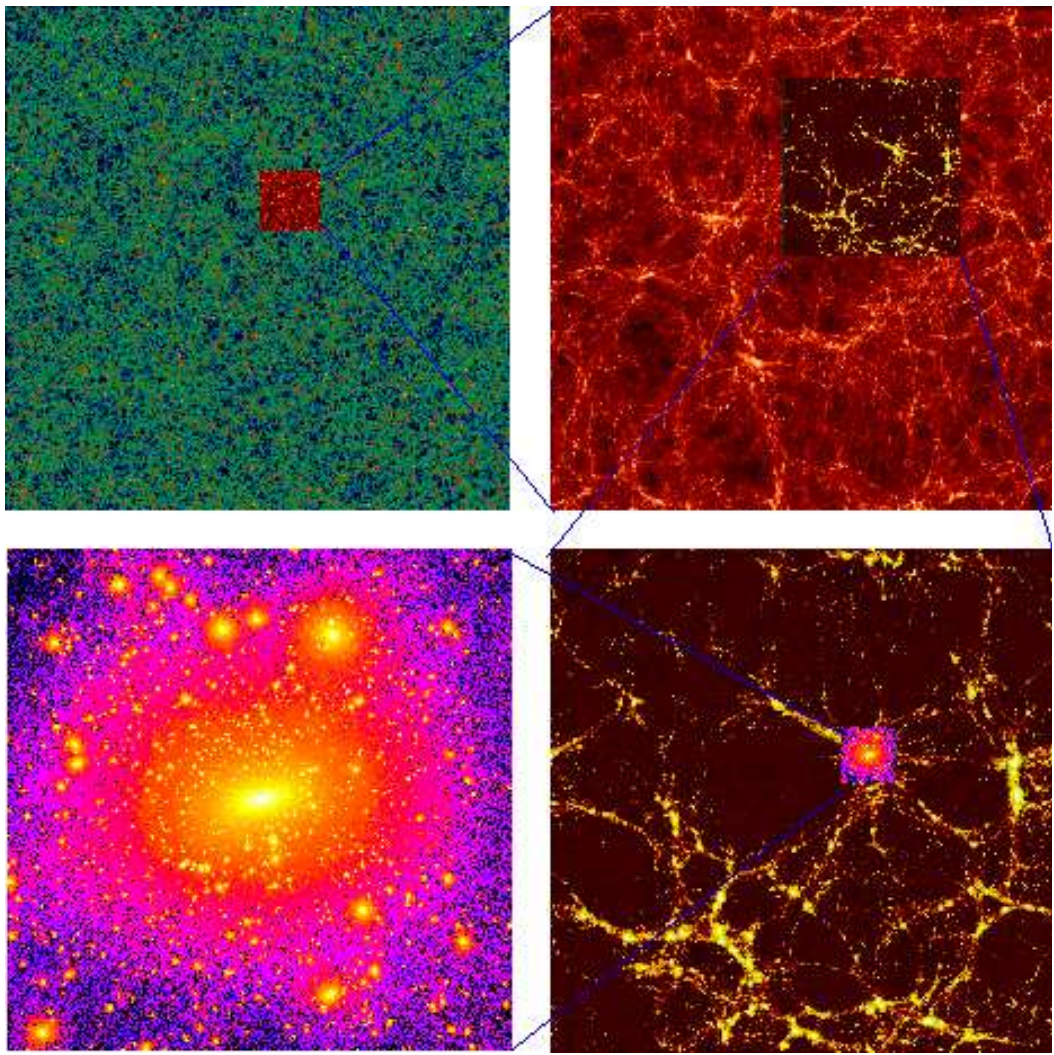,width=1.1\textwidth,,clip=,angle=0,bbllx=74,bblly=256,bburx=446,bbury=628}}
\
\caption{Slices through 4 different simulations of the dark matter
in the `$\Lambda$CDM" cosmology. Denoting the number of particles in each
simulation by $N$, the length of the simulation cube by $L$, the thickness
of the slice by $t$, and the particle mass by $m_p$, the characteristics of
each panel are as follows.  Top-left (the Hubble volume simulation, Evrard
\etal 2002): $N=10^9$, $L=3000$\hmpc, $t=30$\hmpc, $m_p=2.2\times
10^{12}$\msun. Top-right (Jenkins \etal 1998): $N=16.8
\times 10^6$, $L=250$\hmpc, $t=25$\hmpc, $m_p=6.9 \times 10^{10}$\msun.
Bottom-right (Jenkins \etal 1998): $N=16.8 \times 10^6$, $L=140$\hmpc,
$t=14$\hmpc, $m_p=1.4\times 10^{10}$\msun.  Bottom-left (Navarro \etal
2002): $N=7\times 10^6$, $L=0.5$\hmpc, $t=1$\hmpc, $m_p=6.5\times
10^5$\msun.}
\
\label{fig:4panel}
\end{figure}

For simulations like the ones illustrated in Figure~\ref{fig:4panel}, it is
possible to characterize the statistical properties of the dark matter
distribution with very high accuracy. For example, Figure~\ref{fig:LCDMxi}
shows the 2-point correlation function, $\xi(r)$, of the dark matter (a
measure of its clustering strength) in the simulation depicted in the
top-right of Figure~\ref{fig:4panel} (Jenkins \etal 1998). The statistical
error bars in this estimate are actually smaller than the thickness of the
line.  Similarly, higher order clustering statistics, topological measures,
the mass function and clustering of dark matter halos and the
time evolution of these quantities can all be determined very precisely
from these simulations (e.g. Jenkins \etal 2001, Evrard
\etal 2002). In a sense, the problem of the distribution of dark
matter in the $\Lambda$CDM model can be regarded as largely
solved\footnote{However, the innermost structure of halos like those
in the bottom-left of Figure~\ref{fig:4panel} is still a matter of
controversy.}.

\begin{figure}
\centerline{\epsfig{file=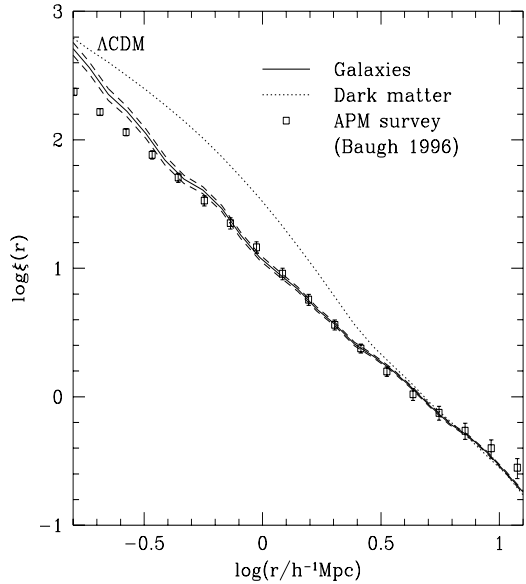,width=0.60\textwidth,height=0.4\textheight,clip=,bbllx=0,bblly=130,bburx=600,bbury=700}}
\caption{Two-point correlation functions. The dotted line shows the
dark matter $\xi_{dm}(r)$ (Jenkins \etal 1998). The solid line shows the
galaxy predictions of Benson \etal (2000), with Poisson errors indicated by
the dashed lines. The points with errorbars show the observed galaxy
$\xi_{gal}(r)$ (Baugh 1996). The galaxy data are discussed in
\S3(b). (Adapted from Benson \etal (2001a).}
\label{fig:LCDMxi}
\end{figure}

In contrast to the clustering of the dark matter, the process of galaxy
formation is still poorly understood.  How then can dark matter simulations
like those of Figure~\ref{fig:4panel} be compared with observational data
which, for the most part, refer to galaxies? On large scales a very
important simplification applies: for Gaussian theories like CDM, it can be
shown that if galaxy formation is a local process, that is, if it depends
only upon local physical conditions (density, temperature, etc), then, on
scales much larger than that associated with individual galaxies, the
galaxies must trace the mass, i.e. on sufficiently large scales,
$\xi_{gal}(r) \propto \xi_{dm}(r)$ (Coles 1993). It suffices therefore to
identify a random subset of the dark matter particles in the simulation to
obtain an accurate prediction for the properties of galaxy clustering on
large scales. This idea (complemented on small scales by an empirical
prescription in the manner described by Cole
\etal 1998) has been used to construct the mock versions of a region of the
APM galaxy survey and of a slice of the 2dFGRS displayed in
Figures~\ref{fig:apm_slice} and~\ref{fig:2df_cone} which also show the
real data for comparison in each case. By eye at least, it is very
difficult to distinguish the mocks from the real data.

\begin{figure}
\centerline{\epsfig{file=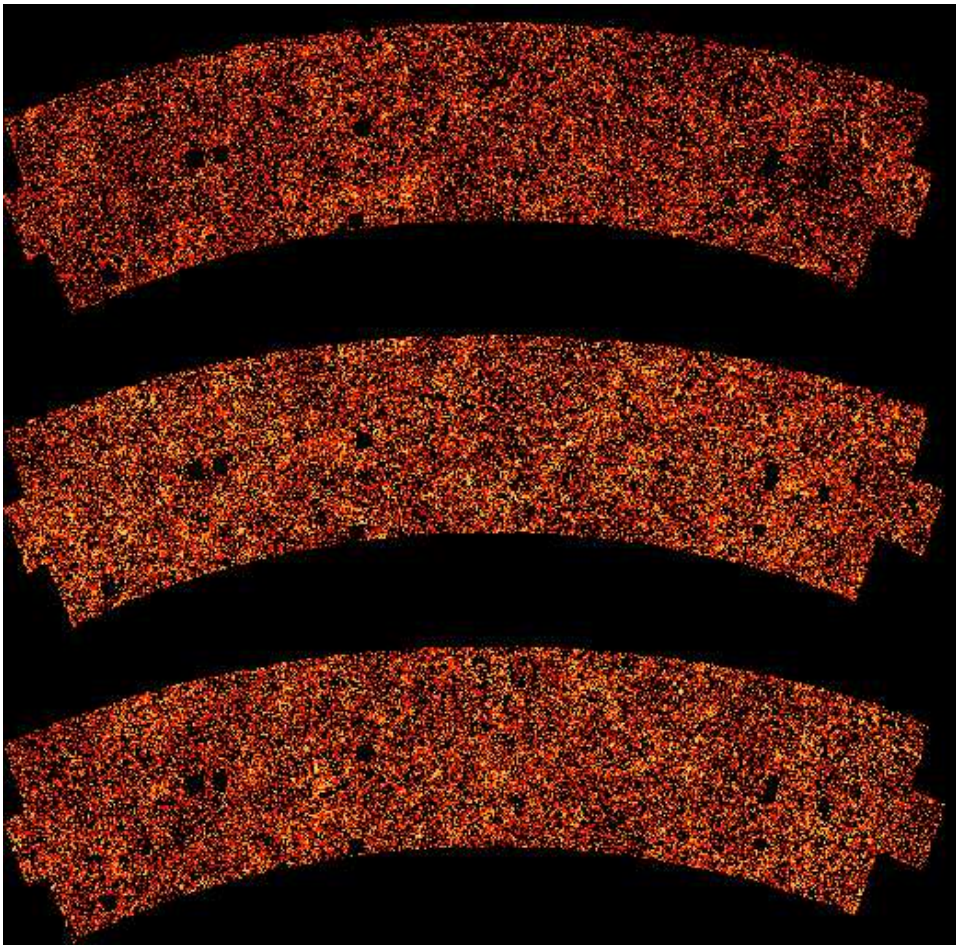,width=1.0\textwidth,clip=,bbllx=168,bblly=293,bburx=427,bbury=549}}
\caption{The region of the APM projected galaxy survey from which the
2dFGRS is 
drawn. Only galaxies brighter than $m_{b_{\rm J}}=19.35$ are plotted. The top
panel is the real data and the other two panels are mock catalogues
constructed from the Hubble volume simulations.}
\label{fig:apm_slice}
\end{figure}

\begin{figure}
\centerline{\epsfig{file=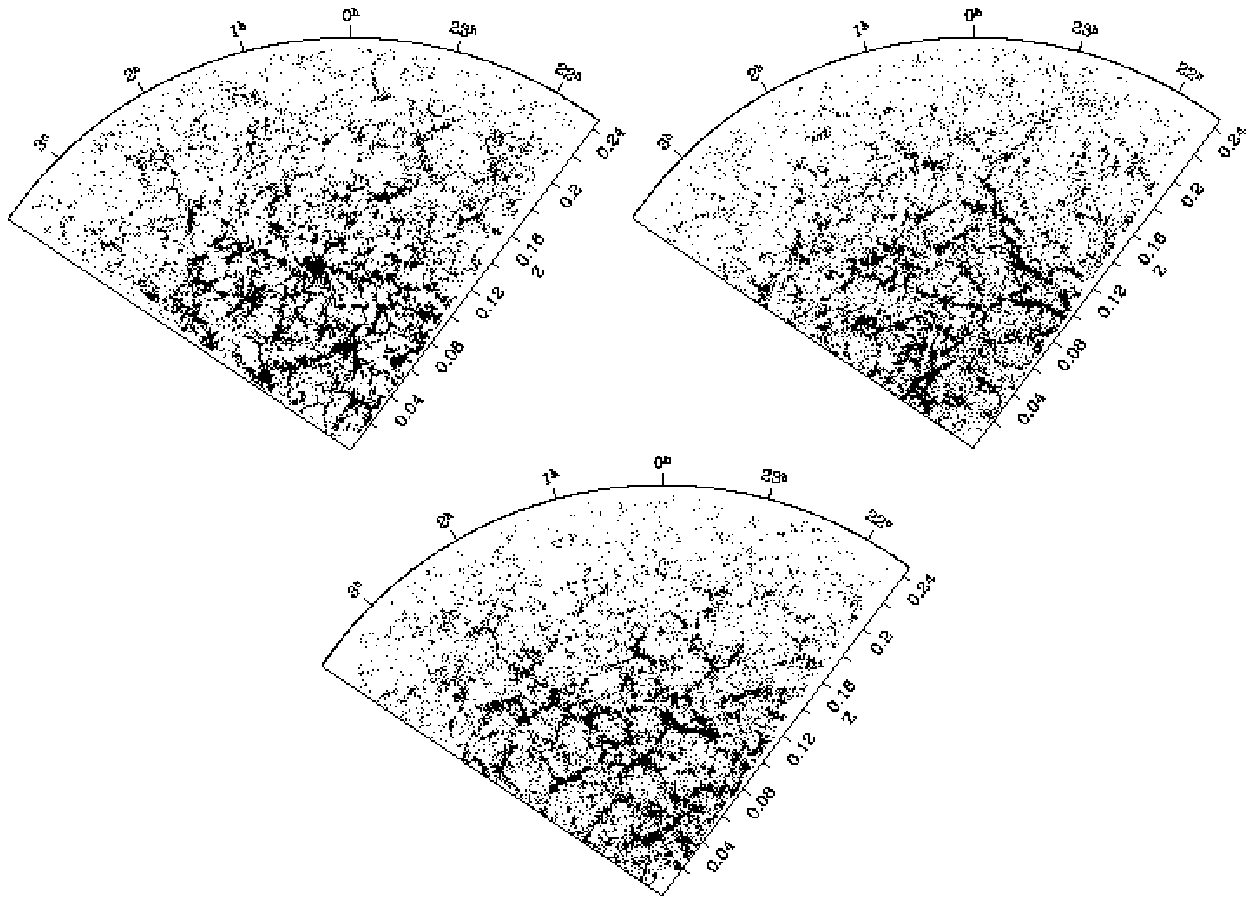,width=1.0\textwidth,clip=,angle=-90,bbllx=145,bblly=223,bburx=451,bbury=619}}
\caption{A 1$^o$ thick slice through the 2dF galaxy redshift survey. The
radial coordinate is redshift and the angular coordinate is right
ascension. The top-left panel is the real data and the other two panels are
mock catalogues constructed from the Hubble volume simulations.}
\label{fig:2df_cone}
\end{figure}

A quantitative comparison between simulations and the real world is carried
out in Figure~\ref{fig:powers}. The symbols show the estimate of the
power spectrum in the 2dFGRS survey (Percival \etal 2001). This is the raw
power spectrum convolved with the survey window function and can be
compared directly with the line showing the theoretical prediction obtained
from the mock catalogues which have exactly the same window function. The
agreement between the data and the $\Lambda$CDM model is remarkably good.

\begin{figure}
\centerline{\epsfig{file=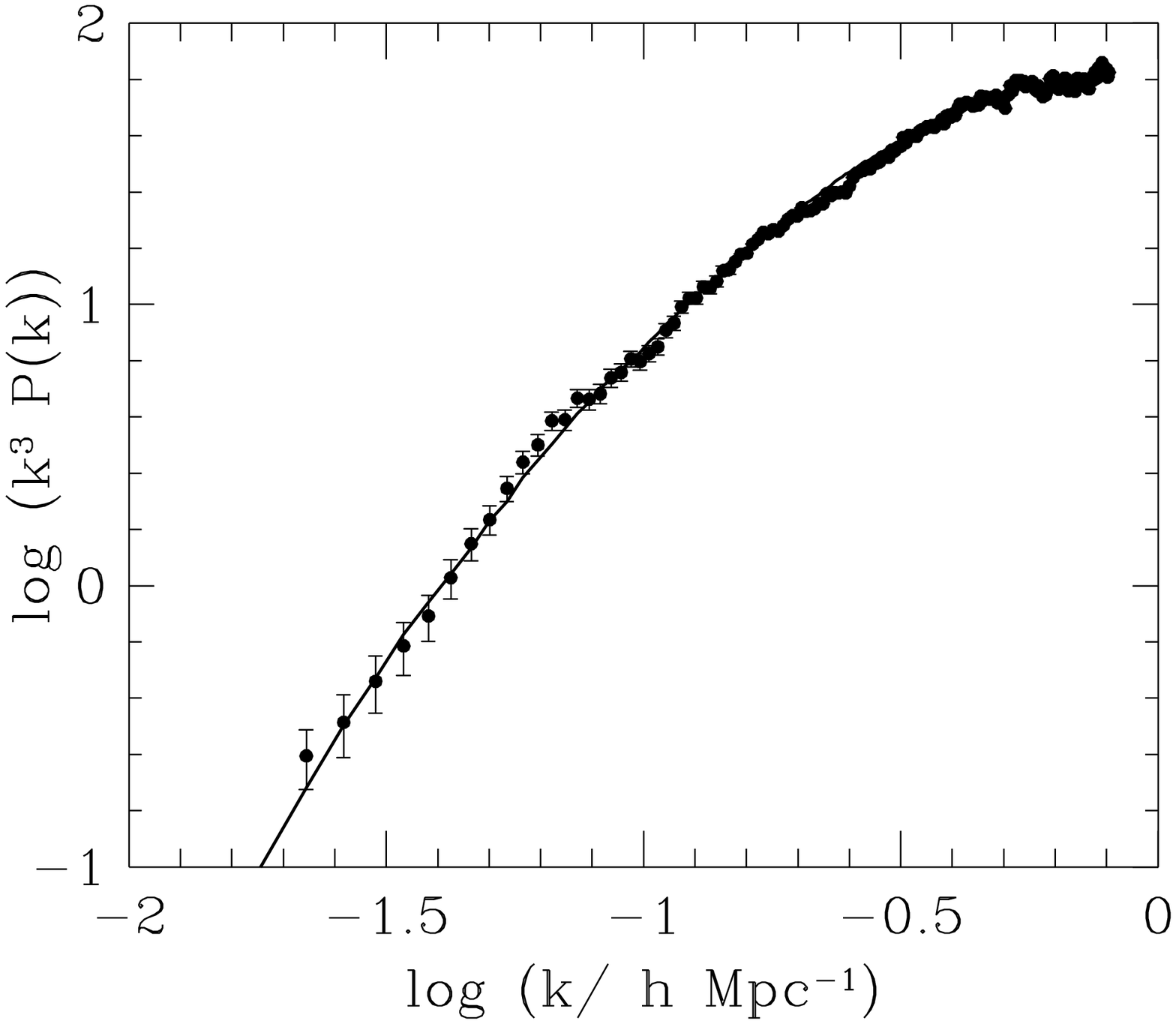,width=0.60\textwidth,clip=,bbllx=0,bblly=0,bburx=550,bbury=520}}
\caption{The power spectrum of the 2dFGRS (symbols) compared with the
power spectrum predicted in the $\Lambda$CDM model (line). Both power
spectra are convolved with the 2dFGRS window function. The model
predictions come from dark matter simulations and assume that, on large 
scales, the distribution of galaxies traces the distribution of
mass. (Adapted from Percival \etal 2001).}
\label{fig:powers}
\end{figure}

\subsection{Galaxy formation} 

Understanding galaxy formation is a much more difficult problem than
understanding the evolution of the dark matter distribution.  In the CDM
theory, galaxies form when gas, initially well mixed with the dark matter,
cools and condenses into emerging dark matter halos.  In addition to
gravity, a non-exhaustive list of the processes that now need to be taken
into account includes: the shock heating and cooling of gas into dark
halos, the formation of stars from cold gas and the evolution of the
resulting stellar population, the feedback processes generated by the
ejection of mass and energy from evolving stars, the production and mixing
of heavy elements, the extinction and reradiation of stellar light by dust
particles, the formation of black holes at the centres of galaxies and the
influence of the associated quasar emission. These processes span an 
enormous range of length and mass scales. For example, the parsec scale
relevant to star formation is a factor of $10^8$ smaller than the scale of
a galaxy supercluster.

The best that can be done with current computing techniques is to
model the evolution of dark matter and gas in a cosmological volume
with resolution comparable to a single galaxy. Subgalactic scales must
then then be regarded as ``subgrid" scales and followed by means of
phenomenological models based either on our current physical
understanding or on observations. In the approach known as
``semi-analytic" modelling (White \& Frenk 1991), even the gas
dynamics is treated phenomenologically using a simple, spherically
symmetric model to describe the accretion and cooling of gas into dark
matter halos. It turns out that this simple model works suprisingly
well as judged by the good agreement with results of full
N-body/gas-dynamical simulations (Benson \etal 2001b, Helly \etal
2002, Yoshida \etal 2002).

The main difficulty encountered in cosmological gas dynamical simulations
arises from the need to suppress a cooling instability present in
hierarchical clustering models like CDM. The building blocks of galaxies
are small clumps that condense at early times. The gas that cools within
them has very high density, reflecting the mean density of the Universe
at that epoch. Since the cooling rate is proportional to the square of the
gas density, in the absence of heat sources, most of the gas would cool in
the highest levels of the mass hierarchy leaving no gas to power star
formation today or even to provide the hot, X-ray emitting plasma detected
in galaxy clusters.  Known heat sources are photoionisation by early
generations of stars and quasars and the injection of energy from
supernovae and active galactic nuclei.  These processes, which
undoubtedly happened in our Universe, belong to the realm of subgrid
physics which cosmological simulations cannot resolve. Different treatments
of this ``feedback" result in different amounts of cool gas and can lead to
very different predictions for the properties of the galaxy population.
This is a fundamental problem that afflicts cosmological simulations even
when they are complemented by the inclusion of semi-analytic techniques. In
this case, the resolution of the calculation can be extended to arbitrarily
small mass halos, perhaps allowing a more realistic treatment of
feedback. Although they are less general than full gasdynamical
simulations, simulations in which the evolution of gas is treated
semi-analytically make experimentation with different prescriptions
relatively simple and efficient (Kauffmann White \& Guiderdoni 1993,
Somerville \& Primack 1999, Cole \etal 2000)

\begin{figure}
\centerline{\epsfig{file=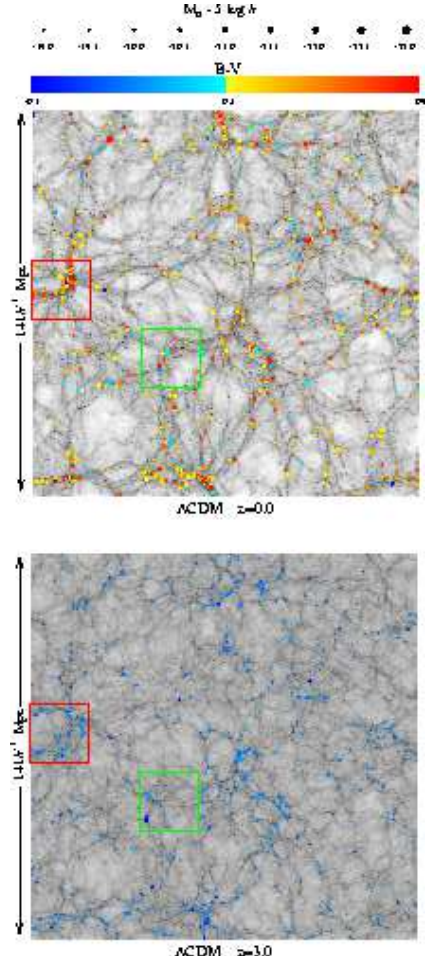,width=1.0\textwidth,clip=,angle=-90,bbllx=55,bblly=326,bburx=467,bbury=506}}
\caption{A slice 10 \hmpcs thick of a simulation of a cubic region 
of side 141 \hmpcs in the $\Lambda$CDM cosmology. The grey scale shows a
slightly smoothed representation of the dark matter in the N-body
simulation. The coloured dots show galaxies; the size of the dots is
proportional to the B-band luminosity of the galaxy and the colour
represents the B-V colour as given on the scale on the top. The top panel
corresponds to redshift $z=0$ and the bottom panel to $z=3$. (Adapted from
Benson \etal 2001a).}
\label{fig:benson}
\end{figure}

The outcome of an N-body dark matter simulation in a $\Lambda$CDM universe
in which the visible properties of the galaxies have been calculated using
the semi-analytic model of Cole \etal (2000) is illustrated in
Fig.~\ref{fig:benson} (Benson \etal 2001a).  Galaxies form mostly 
along the filaments delineated by the dark matter. Red
galaxies predominate in the most massive dark matter halos, just as
observed in real galaxy clusters. This segregation is a natural outcome of
hierarchical clustering from CDM initial conditions. It reflects the fact
that the progenitors of rich clusters form substantially earlier than a
typical dark matter halo of the same mass. Fig.~\ref{fig:lf} shows the
galaxy luminosity function which describes the abundance of galaxies of
different luminosities. The theoretical predictions, shown by the line,
agree remarkably well with the observations but this should not be regarded
as a spectacular success of the theory because the free parameters in the
semi-analytic star formation and feedback model have been tuned to achieve
as good a match as possible to this specific observational dataset. In
particular, the feedback model has been tuned to produce a relatively flat
function at the faint end.

\begin{figure}
\centerline{\epsfig{file=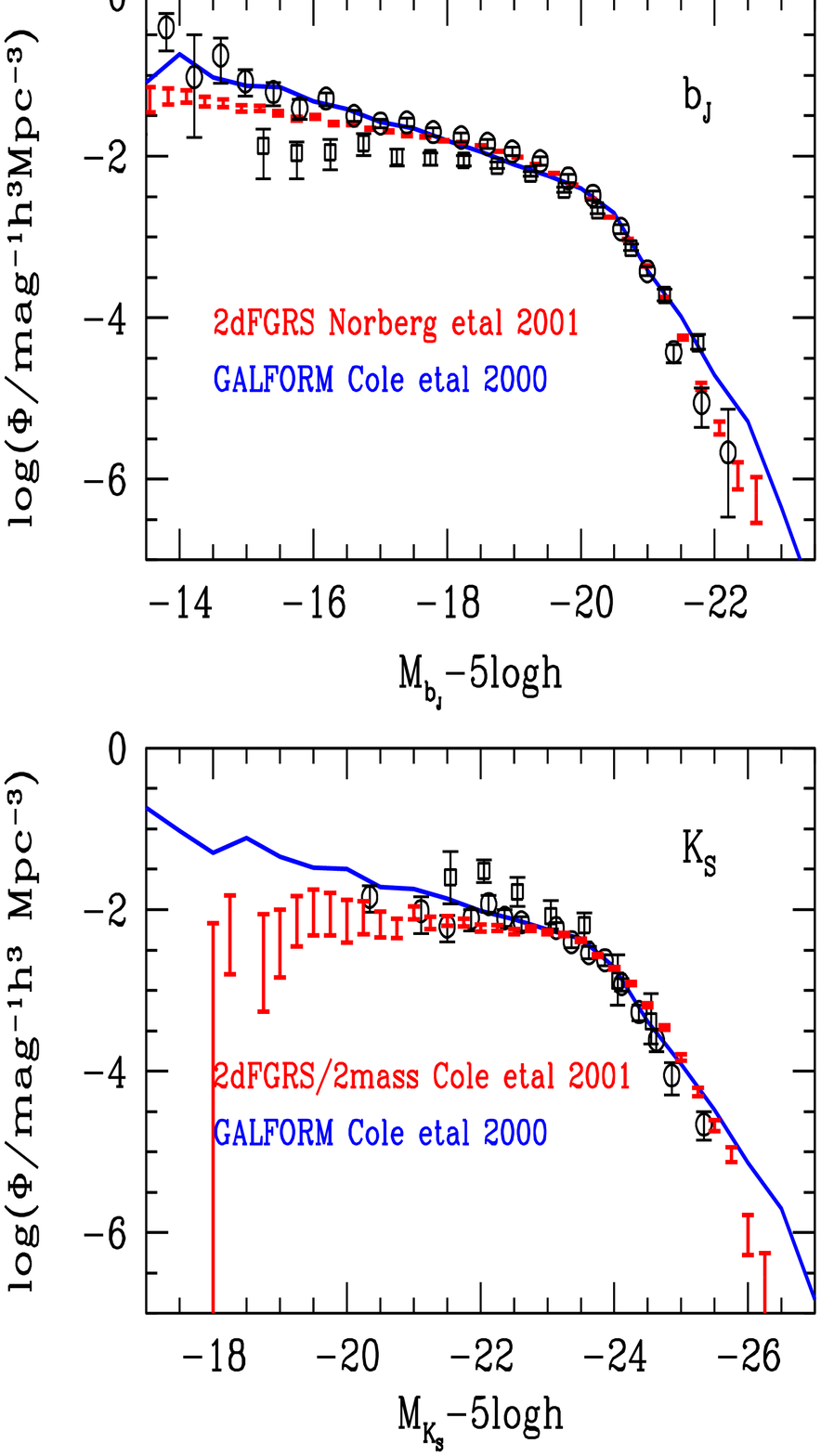,width=0.80\textwidth,clip=,bbllx=60,bblly=400,bburx=560,bbury=810}}
\caption{The galaxy luminosity function. The symbols show the number
of galaxies per unit volume and per unit magnitude measured in various 
surveys, as a function of galaxy magnitude (open circles: Zucca \etal 1997; 
open squares: Loveday \etal 1992; thick error bars: Norberg \etal
2001b). The solid line shows the 
predictions of the semi-analytic model of Cole \etal (2000).}
\label{fig:lf}
\end{figure}

Having fixed the model parameters by reference to a small subset of the
data such as the galaxy luminosity function, we can ask whether the same
model accounts for other basic observational data.  The galaxy
autocorrelation function, $\xi_{gal}(r)$, in the simulations is plotted in
Fig.~\ref{fig:LCDMxi} above. On large scales, it follows $\xi_{dm}(r)$
quite closely, but on small scales it dips below the mass autocorrelation
function. This small scale ``antibias" has also been seen in
N-body/gasdynamical simulations of the $\Lambda$CDM cosmology (Pearce
\etal 1999, 2001, Dave \etal 1999), and in dark matter simulations
that resolve individual galactic halos (Klypin \etal 1999). The galaxy
autocorrelation function in the simulations of Benson \etal (2000)
agrees remarkably well with the observational data (see also Kauffmann
\etal 1999a). This is a genuine success of the theory because no model 
parameters have been adjusted in this comparison. The differences
between the small-scale clustering of galaxies and dark matter result
from the interplay between the clustering of dark matter halos and the
occupation statistics of galaxies in halos which, in turn, are
determined by the physics of galaxy formation. This conclusion, 
discussed in detail by Benson \etal (2000), has led to the development
of an analytic formulation known as the ``halo model" (e.g. Seljak
2000, Peacock \& Smith 2000, Berlind \& Weinberg 2002). 

\begin{figure}
\centerline{\epsfig{file=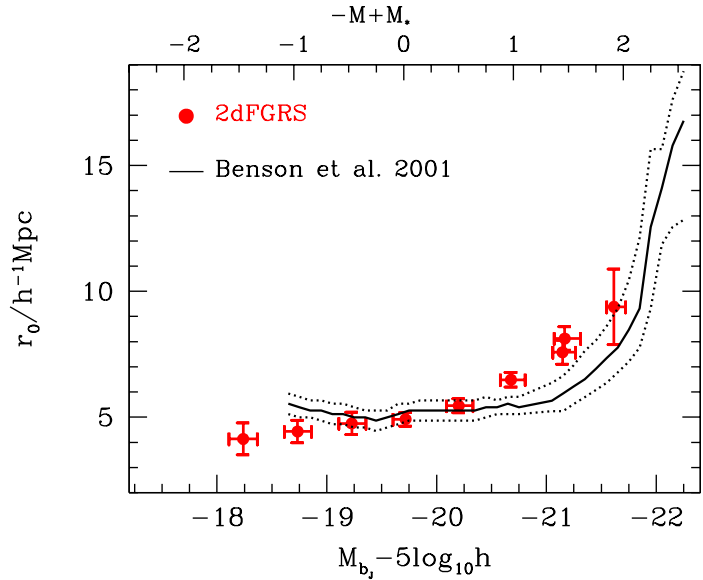,width=0.75\textwidth,clip=,bbllx=40,bblly=230,bburx=565,bbury=665}}
\caption{The correlation length as a function of the luminosity of
different galaxy subsamples. The correlation length is defined as 
the pair separation for which $\xi(r)=1$. The symbols show the results 
from the 2dFGRS and the line the predictions of the simulations of
Benson \etal (2000). (Adapted from Norberg \etal 2001a).}
\label{fig:rovsl}
\end{figure}

Another genuine prediction of the model is the dependence of the strength
of clustering on the luminosity of different subsamples. It can be seen in
Fig.~\ref{fig:benson} that the brightest galaxies are concentrated in the
most massive clusters, leading one to suspect that their autocorrelation
function must be stronger than average. This is indeed the case, as
illustrated in Fig.~\ref{fig:rovsl} which compares the variation of the
clustering length (defined as the pair separation for which $\xi(r)=1$) of
galaxy samples of different intrinsic luminosity in the simulations of
Benson \etal (2001a) with the observational data obtained from the 2dFGRS
by Norberg \etal (2001a). The agreement between theory and observations is
remarkable considering that there are no adjustble parameters in this
comparison. The reason for the strong clustering of bright galaxies is
related to the colour-density relation seen in Fig.~\ref{fig:benson}: the
brightest galaxies form in the highest peaks of the density distribution
which, in initially Gaussian fields, are more strongly clustered than
average peaks which produce less extreme galaxies.

The patch of model universe illustrated in the top panel of
Fig.~\ref{fig:benson} is shown at the earlier epoch corresponding to
redshift $z=3$ (when the universe was only about 20\% of its current
age) in the bottom panel of this figure. The galaxies are now blue,
reflecting the colour of their younger stellar population. There are
fewer galaxies in this plot than in the $z=0$ slice. In fact, this is
the epoch when the first substantial population of bright galaxies
formed in the simulation. As Baugh \etal (1998) argued, the properties
of these model galaxies resemble those of the ``Lyman-break" galaxies
discovered by Steidel \etal (1996), even though different models make
somewhat different predictions for their exact properties (Somerville
\etal 2001). Most models, however, predict that the brightest galaxies
at $z=3$ should be strongly clustered (Kauffmann
\etal 1999b) and, indeed, the models of Baugh \etal (1998) correctly
anticipated that the Lyman-break galaxies would have a clustering
length comparable to that of bright galaxies today (Adelberger \etal
1998). This too should be regarded as a significant success of this
kind of modelling in the $\Lambda$CDM cosmology. As
Fig.~\ref{fig:benson} shows, in contrast to the galaxies, the dark
matter is much more weakly clustered at $z=3$ than at $z=0$,
indicating that galaxies were strongly biased at birth.

\section{Conclusions}

Unlike most computational problems in many areas of science, the
cosmological problem is blessed with known, well-specified initial
conditions. Within a general class of models, it is possible to calculate
the properties of primordial perturbations in the cosmic energy density
generated by quantum processes during an early inflationary epoch. In a
wide family of inflationary models, these perturbations are adiabatic,
scale-invariant and have Gaussian-distributed Fourier amplitudes. The model
also requires an assumption about the nature of the dark matter and the
possibilities have now been narrowed down to non-baryonic candidates of
which cold dark matter particles seem the most promising. An empirical test
of the initial conditions for the formation of structure predicted by the
model is provided by the cosmic microwave background radiation. The tiny
temperature fluctuations it exhibits have exactly the properties expected
in the model. Furthermore, the CMB data can be used to fix some of the key
model parameters such as $\Omega$ and $\Omega_b$, while these data,
combined with other recent datasets such as the 2dFGRS, allow the
determination of many of the remaining parameters such as $\Omega_m$,
$\Omega_{\Lambda}$ and $h$.  It this specificity of the cosmological
problem that has turned simulations into the primary tool for connecting
cosmological theory to astronomical observations.

In addition to well-specified initial conditions, the cosmological dark
matter problem has the advantage that the only physical interaction that is
important is gravity. The problem can thus be posed as a gravitational
N-body problem and approached using the many sophisticated techniques that
have been developed over the past two decades to tackle this
problem. Although on small scales there remain a number of unresolved
issues, it is fair to say that on scales larger than a few megaparsecs, the
distribution of dark matter in CDM models is essentially understood. The
inner structure of dark matter halos, on the other hand, is still a matter
of debate and the mass function of dark matter halos has only been reliably
established by simulations down to masses of order $10^{11}
\Msun$. Resolving these outstanding issues is certainly within reach, but
this will require carefully designed simulations and large amounts of
computing power.

The frontier of the subject at present lies in simulations of the
formation, evolution and structure of galaxies. This problem requires first
of all a treatment of gas dynamics in a cosmological context and a number
of techniques, relying on direct simulations or on semi-analytical
approximations, are being explored. There are quite a few different
approaches to cosmological gasdynamics, but it is reassuring that they all
give similar results in the simplest relevant problem, the evolution of
non-radiative gas during the formation of a galaxy cluster. No detailed
comparisons exist yet for the more complicated case in which the gas is
allowed to cool, but at least one of the gasdynamic simulation techniques,
SPH, gives quite similar results to a simple semi-analytic
approach. Realistic models of galaxy formation, however, will require much
more than a correct treatment of cooling gas. Such models will necessarily
have to include a plethora of astrophysical phenomena such as star
formation, feedback, metal enrichment, etc. The huge disparity between the
submegaparsec scales on which these processes operate and the gigaparsec
scale of the large-scale structure makes it impossible to contemplate a
comprehensive {\it ab initio} calculation. The way forward is clearly
through a hybrid approach combining direct simulation of processes
operating on a limited range of scales with a phenomenological treatment of
the others. There is currently a great deal of activity in the phenomenology
of galaxy formation.

In spite of the uncertainties that remain, all the indications are
that our Universe is well described by a model in which 

\begin{itemize}

\item[(i)] the overall geometry is flat; 

\item[(ii)] the dominant dynamical components are cold dark matter ($\sim
30\%$) and dark energy ($\sim 70\%$) with baryons playing very much a
supporting role ($\sim 4\%$);

\item[(iii)] the initial conditions are quantum fluctuations in the
primordial energy density generated during inflation and

\item[(iv)] structure has grown
primarily as a result of the gravitational instability experienced by mass
fluctuations in an expanding universe.

\end{itemize}

A skeptic is entitled to feel that the current paradigm is odd, to say the
least. Not only is there a need to invoke vast amounts of as yet undetected
non-baryonic cold dark matter, but there is also the need to account for
the dominant presence of a dark energy whose very existence is a mystery
within conventional models of fundamental physics.  Odd as it may seem,
however, this model accounts remarkably well for a large and diverse
collection of empirical facts that span 13 billion years of evolution.

\section{Acknowledgements}

I am grateful to my collaborators for their contribution to the work
reviewed here, especially Carlton Baugh, Andrew Benson, Shaun Cole, Adrian
Jenkins, Cedric Lacey, Peder Norberg, John Peacock, Will Percival, and
Simon White.

\section{References}

\refindent Adelberger K. L., Steidel C. C., Giavalisco M., Dickinson M.,
Pettini M., Kellogg M. (1998), ApJ, 505, 18

\refindent Bardeen, J. M., Bond, J. R., Kaiser, N. and Szalay, A. S. (1986), 
Ap.J., 304, 15 

\refindent Barnes, J., Hut, P. (1986), Nature, 324, 446

\refindent Baugh C. M. (1996), MNRAS, 280, 26

\refindent Baugh C. M., Cole S., Frenk C. S. (1996b), MNRAS, 283, 1361

\refindent Baugh C. M., Cole S., Frenk C. S. Lacey C. G., (1998), ApJ, 498,
504 

\refindent Benson A. J., Cole S., Frenk C. S., Baugh C. M., Lacey C. G.
(2000), MNRAS, 311, 79 

\refindent Benson, A. J., Frenk, C. S., Baugh, C. M., Cole, S., Lacey,
C. G. (2001a), MNRAS, 327, 1041

\refindent Benson, A. J., Pearce, F. R., Frenk, C. S., Baugh, C. M. \&
Jenkins, A. R. (2001b), MNRAS, 320, 261

\refindent Berlind, A. \& Weinberg, D. (2002), ApJ, in press
(astro-ph/0109001) 

\refindent de Bernardis, P. etal (2000), Nature 404, 995

\refindent Borgani, S. etal (2001), ApJ, 561, 13

\refindent Cen, R. (1992), ApJ Suppl 78, 341

\refindent Clowe, D., Luppino, G. A., Kaiser, N., Gioia, I. M. (2000), ApJ, 
539, 540 

\refindent Cole, S., Hatton, S., Weinberg, D. H., Frenk, C. S. (1998),
MNRAS, 300, 945 

\refindent Cole, S., Lacey, C. G., Baugh, C. M., Frenk, C.S. (2000), 
MNRAS, 319, 168

\refindent Coles, P. (1993) MNRAS, 262, 1065

\refindent Colless, M. etal (the 2dFGRS team) (2001), MNRAS, 328, 1039

\refindent Couchman, H. M. P., Thomas, P. A., Pearce, F. R. (1995), ApJ, 
452, 797 

\refindent Dave, R., Hernquist, L., Katz, N. \& Weinberg, D. (1999),
in Proceedings of Rencontres Internationales de l'IGRAP, Clustering at High
Redshift, Marseille 1999, astro-ph/9910221

\refindent  Davis, M., Efstathiou, G., Frenk, C. S. \& White,
S. D. M. (1985), ApJ., 292, 371

\refindent Efstathiou, G., Davis, M., Frenk, C. S. \& White,
S. D. M. (1985), ApJ Suppl, 57, 241

\refindent Efstathiou, G. etal (The 2dFGRS team) (2002), MNRAS, in press, 
(astro-ph/0109152) 

\refindent Efstathiou, G., Sutherland, W. J., Maddox, S. J. (1990), 
Nature, 348, 705

\refindent Eke, V. R., Cole, S. \& Frenk, C. S. (1996), MNRAS, 282, 263

\refindent Eke, V. R., Cole, S.  Frenk, C. S. \& Henry, J. P. (1998), 
MNRAS, 298, 1145 

\refindent Evrard, A. E. (1997), MNRAS, 292, 289.

\refindent Evrard, A.E., MacFarland, T., Couchman, H.M.P., Colberg,  J.M.,  
Yoshida, N., White, S.D.M., Jenkins, A., Frenk, C.S., Pearce, F.R.,
Efstathiou, G., Peacock, J. \& Thomas, P., (The Virgo Consortium) (2002), 
ApJ, in press.

\refindent Fischer, P. etal (2000), AJ 120, 1198

\refindent Frenk, C. S. etal (1999), ApJ, 525, 554

\refindent Freedman, W. L. etal (2001), ApJ, 553, 47

\refindent Gingold, R. A. \& Monaghan, J. J. (1977), MNRAS 181, 375

\refindent Gnedin, N. Y. (1995), ApJ Suppl, 97, 231 

\refindent Guth, A. H. (1981), Phys Rev D, 23, 347

\refindent Hanany, S. etal (2000), ApJ Lett, 545, 5

\refindent Helly, J., Cole, S. M., Frenk, C. S., Baugh, C., Benson, A. \&
Lacey, C. G. (2002), submitted to MNRAS (astro-ph/0202485)

\refindent Jenkins, A., Frenk, C. S., Pearce, F. R., Thomas, P., Colberg, J., 
White, S. D. M., Couchman, H., Peacock, J., Efstathiou, G. \& Nelson,
A. (1998), ApJ, 499, 20

\refindent Jenkins A., Frenk C. S., White S. D. M., Colberg J. M., Cole S.,
Evrard A. E., Yoshida N. (2001), MNRAS, 321,372 

\refindent Kaiser, N. (1987), MNRAS, 227, 1 

\refindent Kauffmann, G., Colberg, J., Diaferio, A., White, S. D. M.
(1999a), MNRAS, 303, 188

\refindent Kauffmann, G., Colberg, J., Diaferio, A., White, S.
D. M. (1999b), MNRAS, 307, 529

\refindent Kauffmann G., White S. D. M., Guiderdoni B. (1993), MNRAS, 264,
201 

\refindent Klypin, A., Gottlober, S., Kravstov, A., Khokhlov, A.M. (1999),
ApJ, 516, 530

\refindent Leitch, E.M. etal (2002), ApJ, 568, 28L 

\refindent Lifshitz, E. M. (1946), J Phys, 10, 116

\refindent Loveday, J., Peterson, B., Efstathiou, G. \& Maddox, S. (1992),
ApJ, 319, 338

\refindent McKay, T. etal (2001) ApJ, in press (astro-ph/0108013)

\refindent Moore, B., Ghigna, S., Governato, F., Lake, G.,  Quinn, T., 
Stadel, J.,  Tozzi, P. (1999), ApJ, 524, 19 

\refindent Navarro, J, F., Power, C., Jenkins, A. R., Frenk, C. S. \&
White, S. D. M. (2002), in preparation

\refindent  Norberg, P., Baugh, C. M., Hawkings, E., Maddox, S., Peacock,
J,. Cole, S., Frenk, C. S., etal (The 2dFGRS team) (2001a), MNRAS, 328,
64 

\refindent  Norberg, P., Cole, S., Baugh, C. M., Frenk, C. S. etal (The
2dFGRS team) (2001b), submitted to MNRAS, (Astro-ph/0111011)


\refindent Peacock, J. A. \& Smith, R. (2000) 318, 1144

\refindent Peacock, J. etal (The 2dFGRS team) (2001), Nature, 410, 169

\refindent Pearce, F. R., Jenkins, A. R., Frenk, C. S., Colberg, J., White, 
S. D. M., Thomas, P., Couchman, H. M. P., Peacock, J. A., Efstathiou,
G. (the Virgo Consortium) (1999), ApJ, 521, 992

\refindent Pearce, F. R., Jenkins, A. R., Frenk, C. S., White, 
S. D. M., Thomas, P., Couchman, H. M. P., Peacock, J. A., Efstathiou,
G. (the Virgo Consortium) (2001), MNRAS, 326, 649

\refindent Peebles, P. J. E. (1970), AJ, 75, 13 

\refindent Peebles, P. J. E. (1980), The large scale structure of the
Universe. (Princeton: Princeton University Press).

\refindent Pen, U-L. (1998), ApJ Suppl, 115, 19 

\refindent Percival, W. etal (the 2dFGRS team) 2001, MNRAS, 327, 1297

\refindent Perlmutter, S. etal (The Supernova Cosmology Project) (1999),
ApJ. 517, 565

\refindent Pierpaoli, E., Scott, D. \& White, M. (2001), MNRAS 325, 77

\refindent Riess, A. etal (1998), AJ, 116, 1009 


\refindent Seljak, U. (2000), MNRAS, 318, 203

\refindent Smoot, G. etal (1992), Apj 396, 1

\refindent Somerville R. S., Primack J. R. (1999), MNRAS, 310, 1087

\refindent Somerville, Rachel S., Primack, Joel R., Faber, S. M. (2001),
MNRAS, 320, 504

\refindent Springel, V., Yoshida, N. \& White, S. D. M. (2001), 
New Astronomy, 6, 79

\refindent Stadel, J. (2000), PhD thesis, University of Washington, Seattle 

\refindent Starobinskii, A. (1982), Phys. Lett. 117B, 175

\refindent Steidel, C. C., Giavalisco, M., Pettini, M., Dickinson, M.,
\& Adelberger, K. L. (1996), ApJ, 462, L17

\refindent Tytler, D., O'Meara, J. M., Suzuki, N. \& Lubin, D. (2000),
Physica Scripta, 85, 12.

\refindent Van Waerbeke, L. etal (2001), A\&A, 374, 757. 

\refindent Viana, P. T. P. \& Liddle, A. (1996), MNRAS, 281, 323 

\refindent Wilson, G., Kaiser, N., Luppino, G., \& Cowie, L. L. (2001),
ApJ, 555, 572  

\refindent White, S. D. M. \& Frenk, C. S. (1991), ApJ, 379, 52

\refindent White, S. D. M., Frenk, C. S. \& Davis, M. (1983) 
ApJ 274, L1. 

\refindent White, S. D. M., Navarro, J. F., Evrard, A. E., Frenk,
C. S. (1993),  Nature, 366, 429. 

\refindent York, D.G, etal (the SDSS collaboration) (2000) AJ, 120, 1579

\refindent Yoshida, N., Stoehr, F., Springel, V. \& White, S. D. M. (2002), 
submitted to MNRAS (Astro-ph/0202341)

\refindent Zucca, E. etal (1997), AA, 326, 477.

\end{document}